\documentstyle[pra,preprint,aps]{revtex}

\begin{document}
\title{Perfectly Translating Lattices on a Cylinder}
\author{Vishal Mehra and Jayme De Luca}
\address{Departamento de F\'{i}sica, Universidade Federal de S\~{a}o Carlos,\\ Rodovia Washington Luiz km 235, 13565-905, Caixa Postal 676 , S\~{a}o\\
Carlos,SP,Brazil}
\maketitle

\begin{abstract}
We perform molecular dynamics simulations on an interacting electron gas
confined to a cylindrical surface and subject to a radial magnetic field and
the field of the positive background. In order to study the system at lowest
energy states that still carry a current, initial configurations are
obtained by a special quenching procedure. We observe the formation of a
steady state in which the entire electron-lattice cycles with a common
uniform velocity. Certain runs show an intermediate instability leading to
lattice rearrangements. A Hall resistance can be defined and depends
linearly on the magnetic field with an anomalous coefficient reflecting the
manybody contributions peculiar to two dimensions.
\end{abstract}

\newpage

\section{\protect\bigskip Introduction}

The advent of modern computers has made it possible to study
many-dimensional dynamical systems in detail and to ferret out previously
unsuspected characteristics. Phenomena such as relaxation to equilibrium 
\cite{deluca}, lack of ergodicity \cite{thirring}, and connection of
dynamical to statistical properties \cite{ruffo} have been explored
numerically. A category of models that continue to enjoy great attention is
that of interacting-electron systems. In this paper we consider a particular
member of this class: the classical dynamics of the confined two-dimensional
electron gas (2DEG). While various aspects of 2DEG have been well-studied,
it remains a convenient and relatively tractable system to understand
dynamical properties of complex systems. It is well-known that at low
temperatures the electron system freezes into an hexagonal Wigner lattice 
\cite{peeters}. Extensive studies have been made of the possible
conformations that result when this simple system is perturbed \cite
{d93,tkt97}. A simple example is provided by the inter-layer coupling
between vertically separated 2DEGs which enhances the stability of the
square lattice relative to the hexagonal \cite{SSP60}. The melting and
structural transitions have also been well-studied \cite{transition}. The
confinement is also a possible perturbation---ideal hexagonal lattice is
obtained only for infinite system.

However, commonly used confining potentials preserve the hexagonal character
with only slight edge distortions. Perturbed 2DEG models have relevance
beyond electron systems: experiments on dusty plasmas \cite{dusty} and ion
plasmas \cite{ion.plasma} have been usefully explained in terms of formation
of few layered and bi-layered Coulomb lattices.

The classical behavior of a confined 2DEG subject to an external magnetic
field has been of some recent theoretical use \cite
{Magnetic1,magnetic2,magnetic3}. In particular the magnetotransport
phenomena in 2DEG have lead to important physical insights. New systems have
been proposed in which the electron motion in 2DEG is nontrivially altered
by a (possibly non-homogeneous) magnetic field \cite{PMI96}. Some
possibilities have been realized by advances in the semiconductor technology 
\cite{skww98}. Such systems are studied quantum mechanically for a full
understanding but many features can already be appreciated within the
classical theory. For example, the classical chaos was found to control the
low-field transport in systems with competing magnetic and electrostatic
modulation \cite{HS99,classical,fgk92}. In this paper we analyze the
structure and dynamics of 2DEG on a novel geometry. The electrons are
constrained to the surface of a cylinder of radius $R_{d}$ and confined
along the $z$ direction to a strip of width $W=2\pi R_{d}$ by the potential
of the uniform static positive background. We chose this geometry because it
allows the electrons to carry a current along the $\phi $ direction. We \
apply the external electric and magnetic fields and observe the
time-dependent response.

We have been motivated to this problem by certain reflections on usual
textbook derivations of the classical 3-dimensional Hall effect \cite
{textbook}. These derivations are based on free-electron Drude theory and
consider independent electrons drifting with a common and constant drift
velocity $V$. It is then easy to show that a constant transverse electric
field, generated by excess boundary charge, can balance the magnetic force
and bring a steady state. In the context of a classical 2-dimensional many
degree-of-freedom interacting system this simple picture could be
dynamically unstable. \ Here we conduct numerical experiments to see if a
regime of interacting electrons drifting with a common uniform velocity can
be attained in a 2D many-body system. A peculiarity of 2DEG is that the
Coulomb field escapes from the surface; in 3D a constant transverse Hall
field can be nicely produced by the boundary electrons only, while in 2D,
because of Gauss's law, these boundary electrons can only produce a $1/r$
field (the field of a charged wire). We see then that a global charge
redistribution is called upon for to produce a constant field.

\section{Simulation and results}

The Lagrangian for $N$ interacting electrons confined by the background
potential $V_{CON}$ and subject to an external radial magnetic field $B$ is 
\begin{equation}
{\cal L}=\sum_{i}\frac{1}{2}m\dot{\vec{x}_{i}}^{2}-\sum_{i>j}\frac{e^{2}}{%
r_{ij}}-V_{CON}(z_{i})-BR_{d}\sum_{i}z_{i}\dot{\phi _{i}},
\end{equation}
where $\vec{x}_{i}$ is the coordinate of the $i$th electron and $r_{ij}$ is
the distance between $i$th and $j$th electrons and $c$ is the speed of
light. The electronic mass and charge are $m$ and $e$ respectively. To
remove any ambiguity we clarify that distances are not calculated along the
surface but are ordinary three-dimensional distances: $%
r_{ij}^{2}=(z_{i}-z_{j})^{2}+2R_{d}(1-\cos (\phi _{i}-\phi _{j}))$. The
rotation symmetry possessed by ${\cal {L}}$ implies the Noether's constant: 
\begin{equation}
l_{z}=\sum_{i}mR_{d}^{2}\dot{\phi _{i}}-BR_{d}\sum_{i}z_{i}.
\end{equation}
The resulting equations of motion are 
\begin{equation}
m\frac{d^{2}\vec{x}_{i}}{dt^{2}}=-\frac{\partial }{\partial \vec{x}_{i}}%
\left( V_{CON}+\sum_{i\neq j}^{N}\frac{e^{2}}{r_{ij}}\right) +\frac{e}{c}%
\vec{B}\times \frac{d\vec{x}_{i}}{dt},  \label{Unscaled}
\end{equation}
where the confining potential $V_{CON}$ \ of the positive background is
taken to be either

\begin{equation}
V_{CON}=-(\ln (z+W/2)+\ln (-z+W/2)),  \label{confine1}
\end{equation}

or

\begin{equation}
V_{CON}=((2z+W)\ln (2z+W)+(-2z+W)\ln (-2z+W)),  \label{confine2}
\end{equation}
the stripe being symmetrical about $z=0$ and extending from $-W/2$ to $W/2$.
The first potential is flat but rises steeply near the edges. The second
potential has more parabolic character and is an approximation to the exact
potential of a positive background: we found analytical solution for the
potential of a positively charged flat rectangle with the same width and
length of our cylindrical background, and the above second potential
approximates this analytic form. The use of potentials with differing
characters helps in distinguishing any peculiar effect of confinement from
overall observations. These potentials are sketched in Fig.(\ref{confine}).

The form of equation (1) is simplified by using scaled units: lengths may be
scaled by the average interelectronic distance $R$: $\vec{x}\rightarrow R%
\vec{x}$ and time is scaled as $dt\rightarrow \omega ^{-1}d\tau $ where $%
\omega ^{2}\equiv e^{2}/mR^{3}$. In these units the equations of motion are

\begin{equation}
\frac{d^{2}\vec{x}_{i}}{d\tau ^{2}}=-\frac{\partial }{\partial \vec{x}_{i}}%
\left( \frac{R}{e^{2}}V_{CON}(R\vec{x}_{i})+\sum_{i\neq j}\frac{1}{r_{ij}}%
\right) +\hat{B}\times \frac{d\vec{x}_{i}}{d\tau },  \label{Scaled}
\end{equation}
where $\hat{B}=n^{-3/4}B/\sqrt{(}mc^{2})$ and $n$ is the two-dimensional
number density of the electron gas. For example at the experimentally
attainable $n=10^{10}$ cm$^{-2}$ the above formula gives $B=0.284~\hat{B}$
Tesla. Because of the scale-invariance of the Coulomb interaction, $\frac{R}{%
e^{2}}V_{CON}(R\vec{x}_{i})$ is independent of $R.$ Equation~(4) has two
components for each particle corresponding to $z$ and $\phi $ motions. We
deliberately add an additionally force along the $\phi $ direction
represented by a small electric field $E_{\phi }<<1$ and a weak damping so
that the final equation for $\phi $ is

\begin{equation}
R_{d}^{2}\ddot{\phi}_{i}=\hat{B}\dot{z}R_{d}+E_{\phi }(1-\frac{R_{d}\dot{\phi%
}_{i}}{V})+R_{d}^{2}\sum_{j\neq i}\frac{\sin (\phi _{i}-\phi _{j})}{%
r_{ij}^{3}}.
\end{equation}
We chose this form of \ damping such that if all electrons cycle with a
constant common velocity $V$, the damping force is balanced by $E_{\phi }.$
The last term in the above equation is just the Coulomb repulsion between
electrons. The equations of motion are integrated numerically using a 7/8
order embedded Runge-Kutta pair with self-adjusted step.

The initial states for the numerical integrations are generated in the
following way: the energy of the confined Coulomb system is 
\begin{equation}
E=\frac{1}{2}\sum_{i}v_{i}^{2}+e^{2}\sum_{i<j}\frac{1}{r_{ij}}%
+\sum_{i}V_{CON}(z_{i}).
\end{equation}
A naive candidate for a minimal-energy initial condition would be one which
minimizes the above energy and with all electrons cycling with the same
velocity $V.$ It is straightforward to find this condition by steepest
descent quenching and one obtains a translating lattice which approximates
an ideal triangular lattice with some edge distortions. The problem with
this condition is that in the presence of the external magnetic field $B$
(which does not appear explicitly in the energy), the translating electrons
deflect upwards causing a lattice deformation. We observe, by integrating
the equations of motion numerically from this condition with inclusion of
the above defined weak dissipation, that the lattice shifts to a different
configuration consistent with the presence of magnetic field and with a
slightly higher steady state energy. (Notice that we are driving the system,
so energy does not have to be constant). \ An equivalent way to accomplish
this same final state is to seek minimum energy configurations with a
certain property: they must describe a uniform motion of the entire system
along the $\phi $ direction under a radial magnetic field $B$ ($v_{i}=V\hat{%
\phi}).$ What we are looking for is that the total force acting on each
electron be zero

\begin{equation}
\nabla \Phi +BV=0,\qquad \Phi =\sum_{i\neq j}\frac{1}{r_{ij}}+V_{CON}.
\end{equation}
This configuration can be obtained by a steepest descent procedure: we
integrate the (modified) quenching equation

\begin{equation}
\frac{d\vec{x}_{i}}{ds}=-\frac{\partial }{\partial x_{i}}(\Phi
+BV\sum_{i}z_{i}/c),
\end{equation}
$s$ being the parameter along the quenching path.

The quenched configurations are obtained for various values of the parameter 
$BV$ and then used as initial conditions for subsequent molecular dynamical
runs. Fig.(2) shows a representative configuration obtained with the
confinement of Eq.(2). The electrons arrange themselves in an hexagonal
lattice slightly distorted by the confinement and magnetic field (notice
that this is a global distortion, not a simple boundary perturbation of a
perfect lattice). Electrons are projected out with an initial $\phi $
velocity $V/R_{d}$ plus small random components along $\phi $ and $z$
directions. Various quantities are calculated along the trajectory: the
instantaneous rotation rate $\sum_{i}\dot{\phi}_{i}/N$ which is related to
the current $I=\sum_{i}\dot{\phi}_{i}/2\pi $; the $\phi $ averaged potential
difference between the top and bottom edges : $V_{H}$ (Hall voltage); and
the Hall resistance $R_{H}=V_{H}/I$. All these quantities are function of
time and hence their time-development is likely to be informative. We also
observe snapshots of the system at regular intervals. These snapshots could
be folded to $0-2\pi $ or left unfolded to preserve information about
angular motion.

We report the results of the simulations of $N=216$ and 484 particles
performed with the confining potential of Eqs (2) and (3). We distinguish
between runs carried out with $E_{\phi }$ zero and nonzero. Simulations with
nonzero $E_{\phi }$ reach a steady state after a brief transient which, is
not attained by zero $E_{\phi }$ runs. We henceforth call this state a
Perfectly Translating Lattice (PTL). In the PTL state the electrons cycle
with a common constant $\phi $ velocity $V/R_{d}$; motion along $z$ being
rapidly damped out. The resulting Hall resistance $R_{H}$ should be a
constant in such circumstances but we observe small fluctuations in $%
R_{H}(t) $. The amplitude of these fluctuations is an irregular function of $%
B$ generally varying between 0.1-1.0 percent. The presence of fluctuating $%
R_{H} $ is not disquieting however and can be understood as an artifact of
the method employed to calculate Hall resistance of small finite number of
electrons: Because we used a finite number of points to average the
potential difference, small instantaneous fluctuations are generated if the
number of electrons is small and we verify that the fluctuations become
smaller for larger $N$ (number of electrons).

Two cases can be further differentiated with regard to the initial phase of
the dynamics. Initial configuration for certain values of the external
magnetic field $B$ turns out to be dynamically unstable. The system makes a
transition to a new configuration through coordinated row-jumping of many
electrons simultaneously. Such a transition is made possible by the
existence of numerous local minima in the energy surface whose presence has
been confirmed numerically by extensive quenching runs starting from
distinct initial conditions. The instability only appears for scaled
magnetic fields greater than 7.0 for $N=216$ and 8.5 for $N=484$ (for
electron density of $10^{10}$ cm$^{-2}$ these fields correspond to 1.9 T and
2.3 T respectively ). In a way, this result shows that the initial state
found by minimization ceased to be stable and another extremum (not the
minimum anymore) of the functional became an stable fixed point, a
bifurcation. This instability is accompanied by slower relaxation to a PTL
state and can be seen in Fig.(3) which plots the spread in the instantaneous
rotation rate vs. time. Even more dramatically, velocity inhomogeneities are
developed if all electrons are released with a common velocity $V$ with no
random components. These inhomogeneities are not long lasting and ultimately
a PTL is attained.

The resultant of this temporarily existing velocity profile can be
visualized most easily in Fig.(4) where the $\phi $ coordinates have not
been folded to $(0-2\pi )$. This striking profile is not apparent after
folding and the finally established PTLs do not differ from PTLs in cases
where this instability is absent. We call this instability a {\sl shearing}
instability since it is visible as a shear in the velocity profile of the
electrons. This shear is seen as a dispersion in unfolded lattice along $%
\phi $ coordinate.

We have carried out runs without forcing and dissipation i.e. $E_{\phi }=0$
and these do not achieve a PTL state. They may or may not display the
initial shearing instability but in all cases the initial randomness in
velocity distribution is magnified and the $z$ dynamics is not damped. An
irregular velocity distribution develops even from a perfectly homogeneous
initial velocity distribution. The instantaneous Hall resistance $R_{H}(t)$
is unsteady with large amplitude fluctuations and the system can not be said
to be in a Hall regime.

 From these simulations a plot of the Hall resistance $R_{H}$ vs. $B$ can be
drawn. Only converged values of $R_{H}$ are used which rules out our
undriven simulations ($E_{\phi }=0$). In Fig.(5) results from unsheared and
sheared states are displayed for $N=216$ and 484. Data from the simulations
employing confining potential of equation~(3) has been plotted for $N=216$
also. All are PTL configurations but the sheared states have experienced the
shearing instability referred to earlier. We observe that $R_{H}$ from the
unsheared states lie on straight lines though different slopes are obtained
for the two confining potentials used: 1.07 and 1.05 for potential (1) and
(2) respectively. Data for $N=216$ and 484 overlap. Points from the sheared
states are scattered haphazardly about this straight line.

How far do these data match our expectations? A plausible argument can be
made for reasonableness of our simulation results: The Hall voltage $V_{H}$
is the difference between the top and the bottom edges of the stripe and may
be expressed as

\begin{equation}
V_{H}=V(W/2)-V(-W/2)=\int_{-W/2}^{W/2}\nabla _{0}\left( \sum_{i}\frac{1}{%
r_{i0}}+V_{CON}\right) .dl_{0},
\end{equation}
where $V=\sum_{i}1/r_{i0}+V_{CON}$ and the integral is taken along a
straight line from bottom to the top edge; $r_{i0}$ is the distance of the $%
i $th electron from integration element. Now the integrand may be split as

\begin{equation}
\nabla _{0}\left( \sum_{i}\frac{1}{r_{i0}}+V_{CON}\right) \approx BV\hat{z}%
+\nabla \left( \frac{1}{r_{n0}}\right) ,
\end{equation}
where $n$ labels the electron nearest to the integration element and $\hat{z}
$ is the unit vector along $z$ direction. This follows from the assuming
that the force-balance condition (Eq.~9) holds in a
neighborhood of the electron nearest to 0:

\begin{equation}
\nabla \left( \sum_{i\neq j}\frac{1}{r_{ij}}+V_{CON}\right) =BV,
\end{equation}
where $i,j$ label electrons. This equation holds only at the position of an
electron and we use as an approximate equality in a neighborhood of the
nearest point $0$. From the above it follows

\begin{equation}
V_{H}=\int BV\hat{z}.dl_{0}+\int \nabla \left( \frac{1}{r_{n0}}\right)
.dl_{0}.
\end{equation}
The first term on the right yields the straight line dependence of $R_{H}$
vs. $B$ with a unit slope but the second term provides a correction. To
evaluate this correction, we take an integration path passing through a
column of electrons, avoiding each electrons by making a small semi-circle
around it. If the distance between the $i$th and $(i+1)$th electrons along
this path is $d_{i}$, this correction evaluates to 
\[
2\sum_{i}\left( \frac{d_{i+1}-d_{i}}{d_{i}}\right) .\frac{1}{d_{i}} 
\]
Hence the correction to the unit slope depends on the degree of compression
that the lattice undergoes under the external magnetic field. A 3D lattice
does not suffer this kind of bulk squishing but only edge distortions. If
the lattice rearrangement is global in the sense that lattice distances are
affected throughout the bulk and not just at the edges then we can fairly
expect a significant alteration of the slope of Hall resistance plot. This
expectation is realized in our simulations as we have seen that the slopes
of $R_{H}$ vs $B$ plot indeed differ from unity by a few percent. Lattices
which are dynamically unstable and undergo rearrangement via the shearing
instability would seem to require more correction according to this picture
and in fact provide a needed check for the theory.

\section{Discussion and Conclusion}

\bigskip In conclusion we have performed dynamical \ simulations on a 2DEG
constrained to a cylindrical stripe and subject to crossed electric and
magnetic fields. The classical confined electron gas has a natural
non-trivial minimal energy state and provides a convenient test bed to study
many-body dynamics and long-range effects. In this work we included
many-body effects on the dynamical picture used in the classical derivations
of the Hall effect in two dimensional systems. We have analyzed the
formation of the steady state presupposed in classical derivations. This
state, which we refer to as a Perfectly Translating Lattice (PTL), in which
all electrons cycle with a common constant velocity, is formed by a
relaxation process. The initial configurations are obtained by a generalized
quenching procedure. This initial configuration is liable to be dynamically
unstable for magnetic fields above a threshold. The quenching procedure
sometimes yields rather shallow local minima which readily allow further
rearrangements to nearby wide basins. The inter-basin like motion is
manifested in simultaneous jumping of many electrons and a slower relaxation
to PTL state. The Hall resistance $R_{H}$ can be calculated and plotted as a
function of external magnetic field $B.$ Appropriately for PTL states $R_{H}$
is a linear function of $B$ except for sheared states. An explanation has
been put forward based upon force-balance condition as it obtains for PTL
states. If one repeated the same study for a confined 3D electron-gas, the
minimal energy state would be a tridimensional Wigner lattice and inclusion
of a drift velocity in the presence of an external magnetic field would
produce only boundary charge rearrangement. The obtained PTL would be the
same ideal Wigner lattice and there would be no many-body correction to the
Hall coefficient. Our results are specific to 2D classical systems.

The present work evolved from our earlier attempts to study the same problem
but with the electrons interacting via the Darwin Lagrangian, which is the
first relativistic correction to the Coulomb interaction. We find that the
relativistic corrections break the scale-invariance of the Coulomb
interaction, even in the absence of a magnetic field, only by requiring a
critical density of the electron gas\cite{Vishal}. Even though these is a
much richer dynamical system, we did not continue the studies because the
equations of motion are algebraic-differential and become impossible to
integrate above the critical density even by use of the modern specialized
integrators RADAU \cite{RADAU}and DASSL \cite{DASSL}.

\begin{figure}[tbp]
\caption{Confining potentials of Eq.\ref{confine1} (full line) and Eq.\ref
{confine2} (dashed line) compared. On horizontal axis is scaled $z$
coordinate $z/R$ and the vertical axis is $V_{CON}$ in units of $e^{2}/R$.}
\label{confine}
\end{figure}

\begin{figure}[tbp]
\caption{Configuration of 2DEG obtained by the special quenching procedure
detailed in the text. This particular state is arrived using confinement of
Eq.\ref{confine1} with $\hat{B}=7.0$ and $V/c=0.05$. The horizontal and
vertical axes denote $\protect\phi $ and $z$ coordinates respectively.}
\label{Pos.ini}
\end{figure}

\begin{figure}[tbp]
\caption{Relaxation to PTL as measured by variance of the instantaneous
rotation rate: $\Delta R=\sum_i(\dot{\protect\phi}_i-\dot{\protect\phi}
>)^{2}$. Three cases are plotted for $N=216$. Slowly-decaying curves are
obtained from $\hat{B}=7$ trajectories which shear in the way discussed in
the text; they differ in the distribution of initial velocities: all
electrons are started with $\dot{\protect\phi}_i=V/R_d,\dot{z}_i=0$ (dotted
curve) or small random components are added to $\dot{\protect\phi}_i=V/R_d,%
\dot{z}_i=0$ (full curve). The nondispersing trajectory (dashed curve) comes
from a $\hat{B}=9$ run and exhibits rapid decay. These behaviors are typical
for dispersing and nondispersing trajectories respectively.}
\end{figure}

\begin{figure}[tbp]
\caption{The unfolded view of a sheared $N=484$ run with confining potential 
\ref{confine1} at $\hat{B}=13.5$. Unfolding preserves some information about
the past angular velocities by displaying the net angular motion. Individual
electrons are marked with pluses. }
\label{Unfold}
\end{figure}
\begin{figure}[tbp]
\caption{The Hall resistance $R_H$ calculated from simulations as detailed
in the text. $R_H$ from unsheared $N=216$ trajectories using $V_{CON}$ of Eq.%
\ref{confine1} are shown with crosses; sheared ones with pluses. Points from 
$N=484$ using potential \ref{confine1} are marked as boxes; these runs were
unsheared for $\hat{B}<8.5$ and sheared for $\hat{B}>8.5$. Data from runs
using $V_{CON}$ of Eq.\ref{confine2} is depicted with stars ($N=216$).The
remnant fluctuations are much smaller than the symbol-size. Hall resistance $%
R_H$ is in units of $e/(\protect\omega R)$ and the $x$-axis is the scaled
magnetic field $\hat{B}=n^{-3/4}B/(mc^{2})$.}
\label{Hall.curve}
\end{figure}

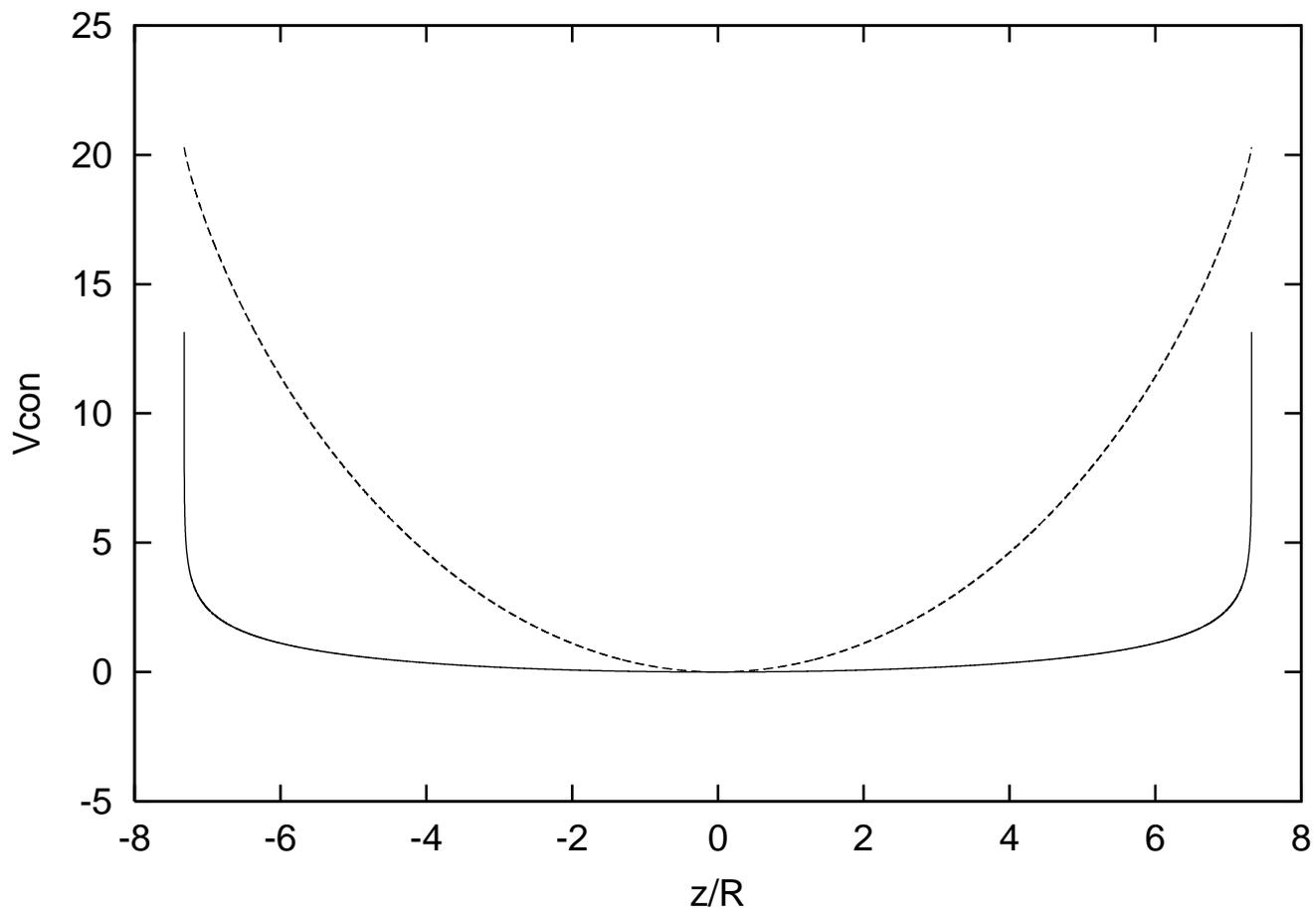
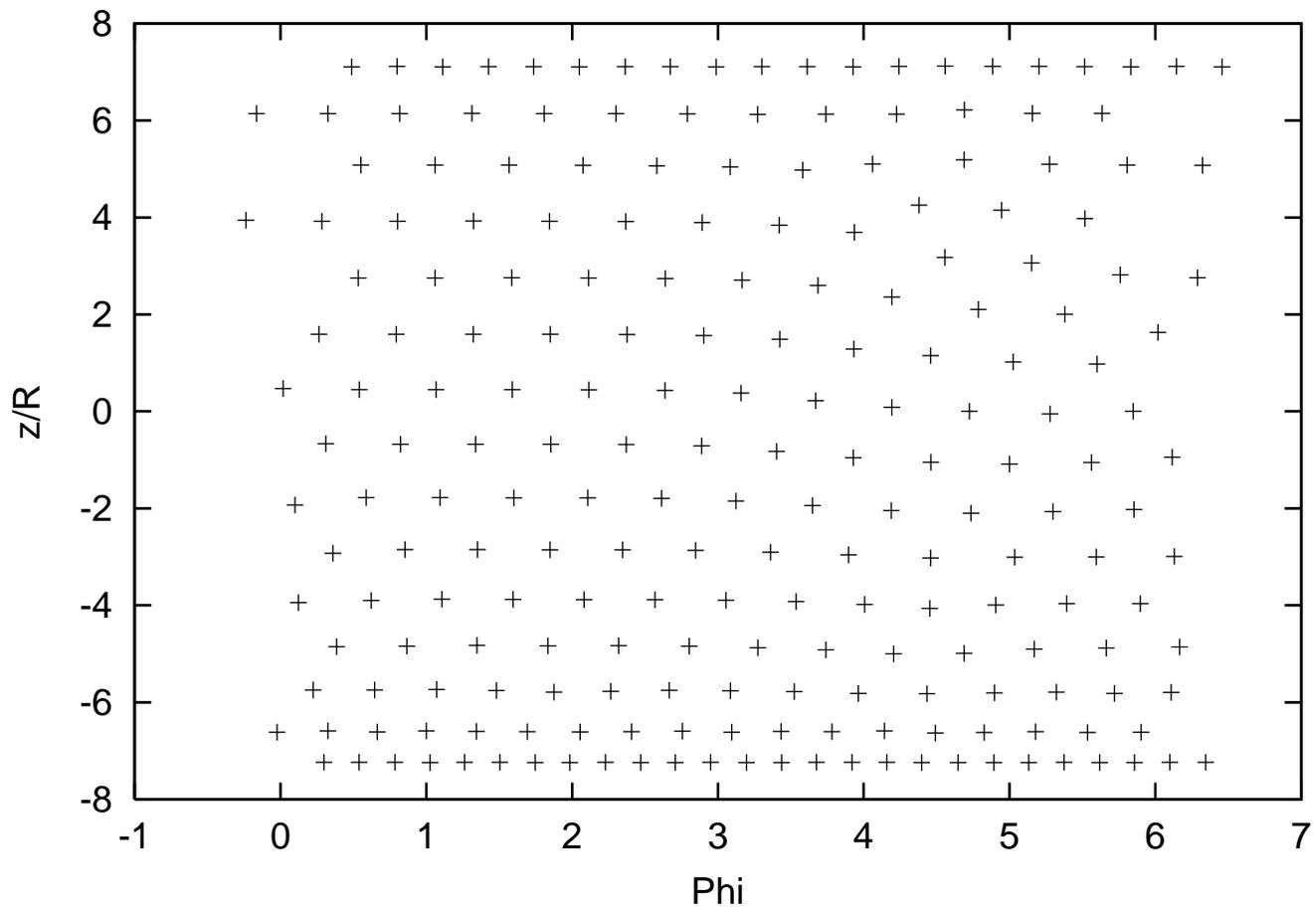
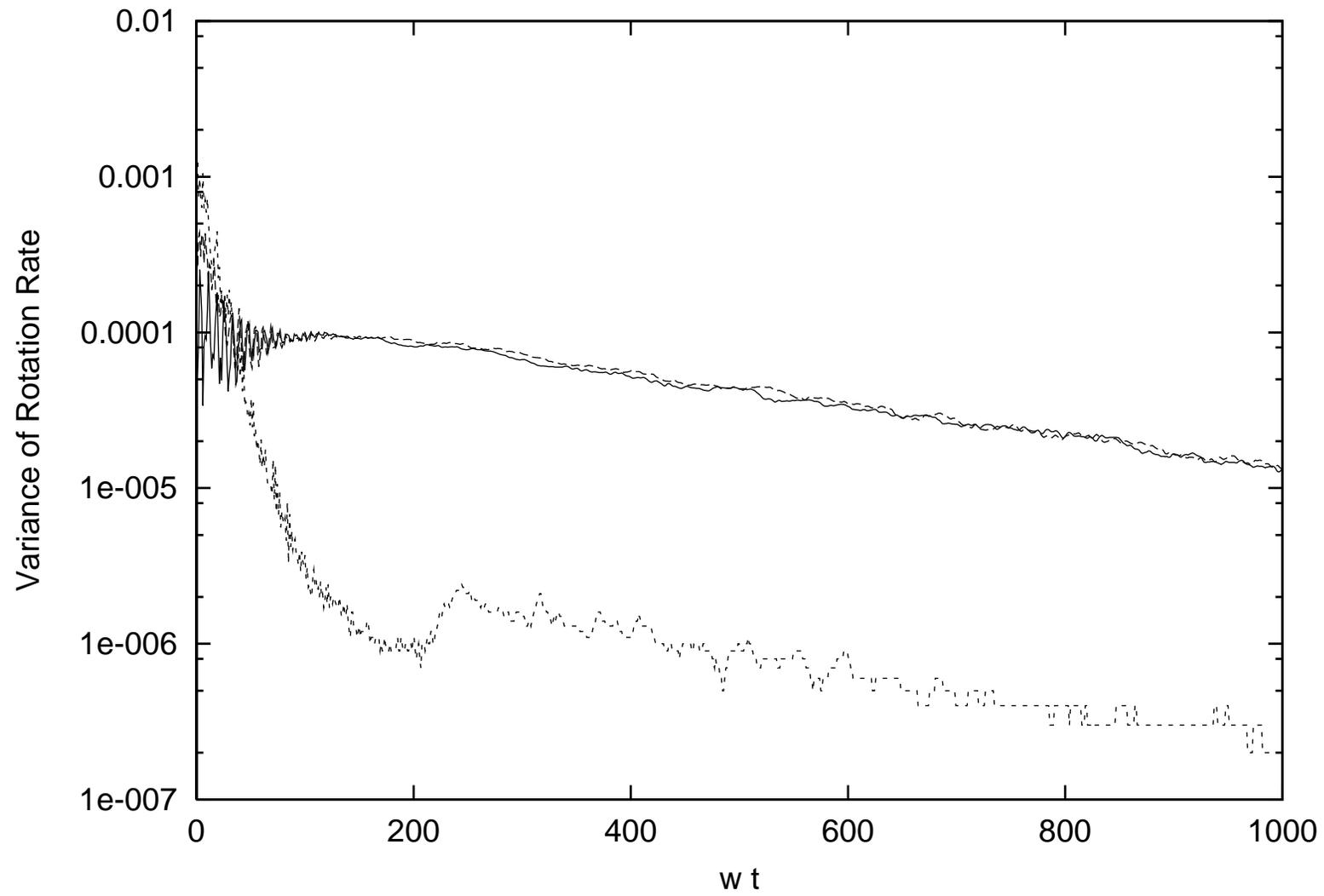
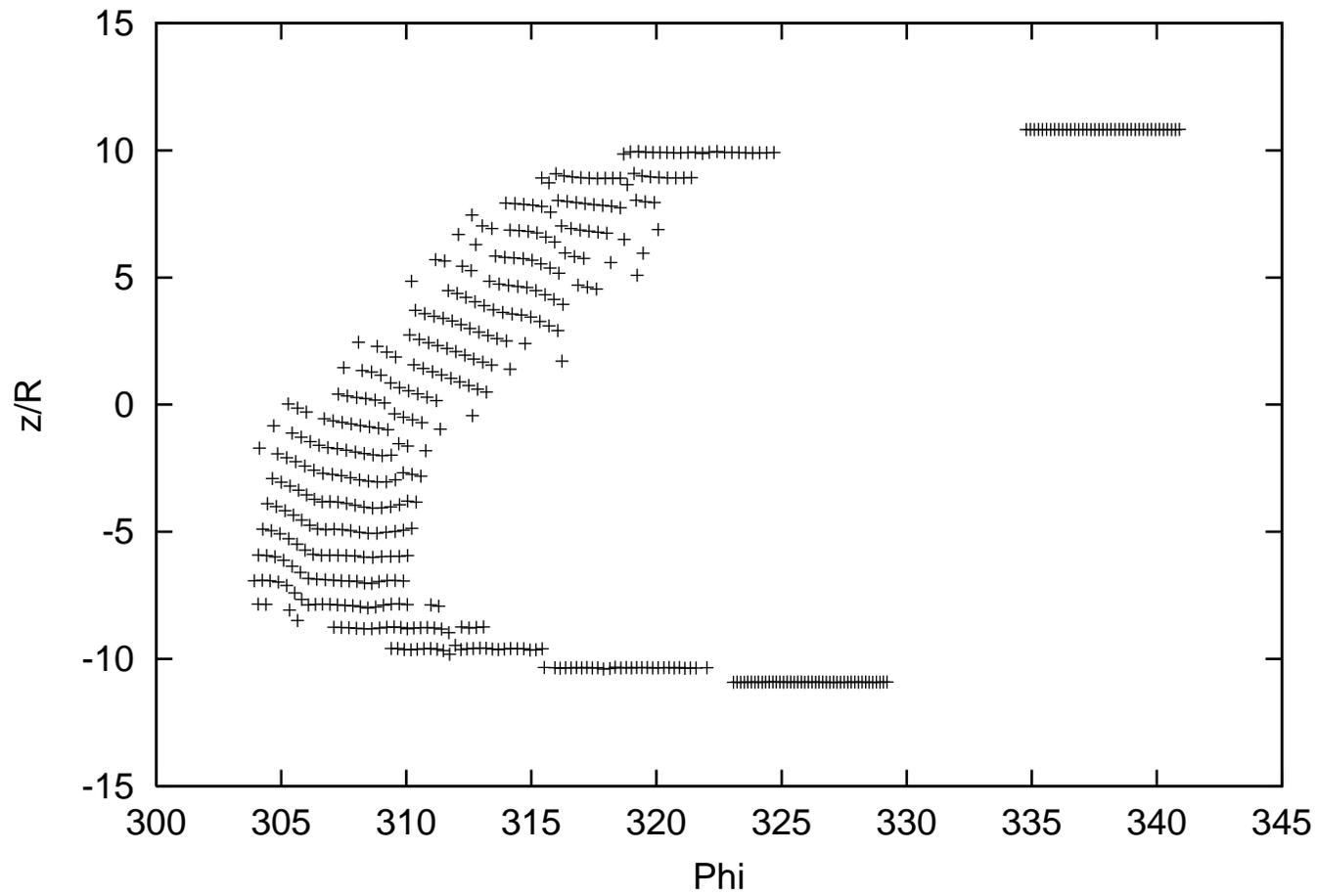
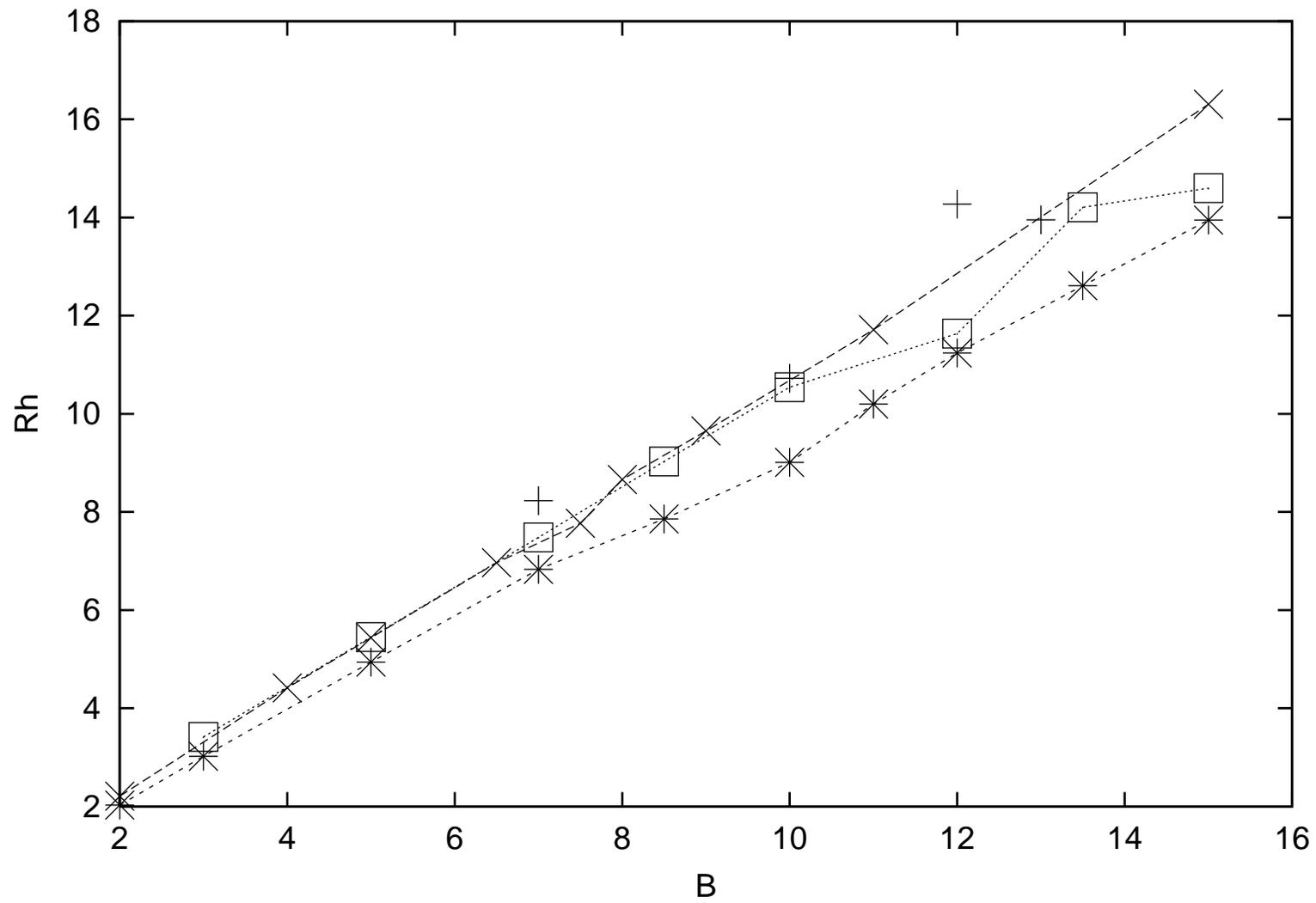
\end{document}